\documentclass[preprint,preprintnumbers,amsmath,amssymb]{revtex4}

\usepackage{graphicx}
\usepackage{dcolumn}
\usepackage{bm}

\usepackage[normalem]{ulem}
\usepackage[dvips]{color}

\renewcommand\sout{\bgroup \color{red} \ULdepth=-.5ex \ULset}

\begin{document}


\title{Isospin dependent global neutron-nucleus optical model potential}

\author{
Xiao-Hua Li$^{1,2}$\footnote{li\_xiaohua@sjtu.edu.cn} , Lie-Wen Chen
$^{1,3}$\footnote{lwchen@sjtu.edu.cn} }
 \affiliation{
1. Department of Physics, Shanghai Jiao Tong University, Shanghai 200240, China\\
2. School of Nuclear Science and Technology, University of South
China, Hengyang, Hunan 421001, China \\
3. Center of Theoretical Nuclear Physics, National Laboratory of
Heavy Ion Accelerator, Lanzhou 730000, China}


\date{\today}

\begin{abstract}
In this paper, we construct a new phenomenological isospin dependent
global neutron-nucleus optical model potential. Based on the
existing experimental data of elastic scattering angular
distributions for neutron as projectile, we obtain a set of the
isospin dependent global neutron-nucleus optical model potential
parameters, which can basically reproduce the experimental data for
target nuclei from $^{24}$Mg to $^{242}$Pu with the energy region up
to 200 MeV.
\end{abstract}

\maketitle

\section{Introduction}

The optical model(OM) is of fundamental importance on many aspects
of nuclear physics~\cite{PGYONG199841}. It is the basis and starting
point for many nuclear model calculations and also is one of the
most important theoretical approaches in nuclear data evaluations
and analyses. The optical model potential (OMP) parameters are the
key to reproduce the experimental data, such as reaction cross
sections, elastic scattering angle distributions, and so on.

Over the past years, a number of excellent local and global
optical potentials for nucleons have been
proposed~\cite{NPA7132312003}\cite{PR18211901969}\cite{PRT201571991}.
Koning and Delarche~\cite{NPA7132312003} constructed a set of
global phenomenological nucleon-nucleus optical model potential
parameters (KD OMP), which can perfectly reproduce the
experimental data for the region of targets from $^{24}$Mg to
$^{209}$Bi with the incident energy from 1 keV to 200 MeV; Weppner
\textit{et al}~\cite{PRC800346082009} obtained a set of isospin
dependent global nucleon-nucleus optical model potential
parameters (WP OMP) with target nuclei region from carbon to
nickel and the projectile energy from 30 to 160 MeV;  Han
\textit{et al}~\cite{PRC810246162010} also obtained a new set of
global phenomenological optical model potential parameters for
nucleon-actinide reactions with energies up to 300 MeV. In the
nucleon optical model potential, the isospin degree of freedom may
play an important role to more accurately describe the
experimental data~\cite{NP356761962}~\cite{AOP1242081980}.
Information on the isospin dependence of the nucleon optical model
potential has been shown to be very useful to understand the
nuclear symmetry
energy~\cite{PRB1347221964,PLB421631972,NPA86512011,PRC820546072010}
which encodes the energy related to the neutron-proton asymmetry
in the equation of state of isospin asymmetric nuclear matter and
is a key quantity for many issues in nuclear physics and
astrophysics (See, e.g., Ref.~\cite{PR4641132008}). On the other
hand, to study the systematics of neutron scattering cross
sections on various nuclei for neutron energies up to several
hundred MeV is a very interesting and important topic due to the
concept of an accelerator driven subcritical (ADS) system in which
neutrons are produced by bombarding a heavy element target with a
high energy proton beam of typically above 1.0 GeV with a current
of 10 mA and the ADS system serves a dual purpose of energy
multiplication and waste incineration (See, e.g.,
Ref.~\cite{nuclth0409005}). Therefore, to construct a more
accurate neutron-nucleus optical model potential is of crucial
importance. The motivation of the present paper is to construct a
new isospin dependent neutron-nucleus optical model potential,
which can reproduce the experimental data for a wider range of
target nucleus than the formers.

This paper is arranged as follows. In Sec. II, we provide a
description of the optical model and the form of the isospin
dependent neutron-nucleus optical potential. Section III presents
the results, and section IV is devoted to the discussion. Finally, a
summary is given in Sec. V.

\section{OPTICAL MODEL AND THE FORM OF  the isospin
dependent neutron-nucleus OPTICAL POTENTIAL}

The phenomenological OMP for neutron-nucleus reaction $V(r,E)$ is
usually defined as follows: \label{eq:whole1}
\begin{equation}
V(r,E)=-V_{v} \,f_r(r)-i \,W_v \,f_v(r)+i \,4 \,a_s\,W_s
\,\frac{df_s(r)}{dr}+\lambda\!\!\!{^-}_\pi^2
\,\frac{V_{so}+iW_{so}}{r} \,\frac{df_{so}(r)}{dr}
\,2\vec{S}\cdot\vec{l} ,\label{subeq:1}
\end{equation}
where $V_v$ and $V_{so}$ are the depth of real part of central
potential and spin-orbit potential, respectively; $W_v$, $W_s$ and
$W_{so}$ are the depth of imaginary part of volume absorption
potential, surface absorption potential and spin-orbit potential,
respectively. The $f_{i}$ ($i=v,s,so$) are the standard Wood-Saxon
shape form factors.

In this work, according to Lane Model~\cite{NP356761962}, we add the
isospin dependent terms in the $V_v$, $W_v$ and $W_s$, which can be
parameterized as:
\begin{eqnarray}
V_v=V_0+V_1 \,E +V_2 \,E^2+(V_3+V_{3L}\,E) \,(N-Z)/A,
\end{eqnarray}
\begin{eqnarray}
W_s=W_{s0}+W_{s1} \,E+(W_{s2}+W_{s2L}\,E) \,(N-Z)/A
\end{eqnarray}
\begin{eqnarray}
W_v=W_{v0}+W_{v1} \,E+W_{v2} \,E^2+(W_{v3}+W_{v3L}\,E)\,(N-Z)/A
\end{eqnarray}

The shape form factors $f_{i}$ can be expressed as
\begin{eqnarray}
f_i(r)=[1+\exp((r-r_i \,A^{1/3})/a_i)]^{-1}\;\; with\;\; i=r,v,s,so
\label{subeq:2}
\end{eqnarray}
where
\begin{eqnarray}
r_i=r_{i0}+r_{i1} \,A^{-1/3} \qquad \qquad \qquad with \quad
i=r,v,s,so
\end{eqnarray}
\begin{eqnarray}
a_i=a_{i0}+a_{i1} \,A^{1/3} \qquad \qquad \qquad with \quad
i=r,v,s,so
\end{eqnarray}
In  above equations, $A = Z + N$ with $Z$ and $N$ being the number
of protons and neutrons of the target nucleus, respectively; $E$ is
the incident neutron energy in the laboratory frame;
$\lambda\!\!\!{^-}_\pi^2$ is the Compton wave length of pion, and
usually we use $\lambda\!\!\!{^-}_\pi^2 = 2.0$ fm${^2}$.

{\small APMN} \cite{NuclSciEng1411782002} is a code to automatically
search for a set of optical potential parameters with smallest $\chi
^{2}$ in $E \leq 300$ MeV energy region by means of the improved
steepest descent algorithm~\cite{SDalgorithm}, which is suitable for
non-fissile medium-heavy nuclei with the light projectiles, such as
neutron, proton, deuteron, triton, $^{3}$He, and $\alpha$. The
optical potential in {\small APMN}~\cite{NuclSciEng1411782002} has
been modified based on the standard BG form~\cite{PR18211901969},
i.e. Woods-Saxon form for the real part potential $V_v$ and the
imaginary part potential of volume absorption $W_{v}$; derivative
Woods-Saxon form for the imaginary part potential of surface
absorption $W_{s}$; and Thomas form for the spin-orbital coupling
potential $V_{so}$ and $W_{so}$. It should be noted that all the
radius and diffusiveness parameters in the standard BG optical
potential form are constant, not varying with the mass of target
nuclei. In the present work, they are modified as functions of the
mass of target nuclei according to our former
work~\cite{NPA7891032007}. We modify the {\small APMN} code
according to the the present form of the isospin dependent global
neutron-nucleus optical model potential and thus totally 32
adjustable parameters are involved in the code {\small
APMN}~\cite{NuclSciEng1411782002}.

In the code {\small APMN}~\cite{NuclSciEng1411782002}, the
compound nucleus elastic scattering is calculated with the
Hauser-Feshbach statistic theory with Lane-Lynn width fluctuation
correction~\cite{Lane-Lynn1957}(WHF), which is designed for
medium-heavy target nuclei. For these nuclei, the spaces between
levels are usually small, the concepts of continuous levels and
level density can be properly used for description of higher
levels, say, their excited energies are higher than the combined
energy of the emitting particle in compound nucleus. In the code
{\small APMN}, the Hauser-Feshbach theory supposed that after the
compound nucleus emits one of the six particles--n, p, d, t,
$\alpha$ and $^3$He, or a $\gamma$ photon, all discrete levels of
the residual nucleus de-excite only through emission of $\gamma$
photons, not permitting emission of any particles. For
medium-heavy target nuclei, when the incident energy increase to
about 5--7 MeV, the cross sections of the compound nucleus elastic
scattering usually will drop to very small values in comparison
with the shape elastic scattering; so there is no need for
considering pre-equilibrium particle emission.

\section{ RESULTS }

Our theoretical calculation is carried out within the
non-relativistic frame and the relativistic kinetics corrections
have been neglected because they are usually very small when the
projectile energy $E \leq 200$ MeV (See, e.g.,
Ref.~\cite{PRC730546052006}). In the present work, we choose the
existing experimental data of neutron elastic scattering angular
distributions with the incident energy region from $0.134$ to $225$
MeV for the $45$ target nuclei shown in Table I as the data base,
for searching for global neutron optical potential parameters. These
data shown in Table I have been also used in the work of Koning and
Delarche~\cite{NPA7132312003}. In this work, all of experimental
data used are taken from EXFOR (web address:
http://www.nndc.bnl.gov/). As for the data error, we take the values
given in EXFOR if they are available (we note here that more than
$90\%$ data considered in the present work have data error in
EXFOR); in the case that the data errors are not provided in EXFOR,
we take them as $10\%$ of the corresponding experimental data, which
roughly corresponds to the mean value of the available experimental
data error.

We use the global neutron optical model potential parameters of
Becchetti and Greenless~\cite{PR18211901969} as starting point. The
value of zero has been used as the initial values for the parameters
that we add newly in the code {\small APMN}.

Through the calculation of {\small APMN} code, we obtain a new set
of isospin dependent global neutron-nucleus optical model potential
parameters which can be expressed as following:

\begin{eqnarray}
 V_v=54.983-0.3278E+0.00031
E^{2}-(18.495-0.219E)(N-Z)/A ~(\textrm{MeV})
\end{eqnarray}
\begin{eqnarray}
 W_s=11.846-0.182E-(16.66-0.0141E)(N-Z)/A ~(\textrm{MeV})
\end{eqnarray}
\begin{eqnarray}
W_v=-2.5028+0.2144E-0.00126E^{2}-(0.000248-0.2139E)(N-Z)/A
~(\textrm{MeV})
\end{eqnarray}
\begin{eqnarray}
&&a_r=0.696-0.00064A^{1/3} ~(\textrm{fm}),\quad  a_s=0.563-0.0137A^{1/3} ~(\textrm{fm})\\
&&a_v=0.912+0.0539A^{1/3} ~(\textrm{fm}),\quad  a_{so}=0.677+0.0203A^{1/3} ~(\textrm{fm})\\
&&r_r=1.173-0.002A^{-1/3} ~(\textrm{fm}),\quad  r_s=1.278-0.014A^{-1/3} ~(\textrm{fm})\\
&&r_v=1.266+0.02A^{-1/3} ~(\textrm{fm}),\quad  r_{so}=0.828+0.01A^{-1/3} ~(\textrm{fm}) \\
&&V_{so}=8.797 ~(\textrm{MeV}),\quad W_{so}=0.019 ~(\textrm{MeV})
\end{eqnarray}
where the unit of the incident neutron energy $E$ is MeV.

With above optical model potential parameters, we calculate the
angular distributions of elastic scattering for many nuclei with
neutron as projectile. Some of the calculated results and
experimental data of elastic scattering angular distributions are
shown in Fig. 1 to Fig. 12 where the corresponding results from KD
OMP are also included for comparison.

\section{DISCUSSION  }

The $\chi^{2}$ represents the deviation of the calculated values
from the experimental data, and in this work it is defined as
follows: \label{eq:whole3}
\begin{equation}
\chi^2=\frac{1}{N}\sum \limits_{n=1}^{N}\chi_n^2,
\end{equation}
with
\begin{eqnarray}
\chi_n^2=\frac{1}{N_{n,el}}\sum
\limits_{i=1}^{N_{n,el}}\frac{1}{N_{n,i}}\sum
\limits_{j=1}^{N_{n,i}}(\frac{\sigma_{el}^{th}(i,j)-\sigma_{el}^{exp}(i,j)}{\Delta
\sigma_{el}^{exp}(i,j)})^2,
\end{eqnarray}
where $\chi_n^{2}$ is for a single nucleus, and $n$ is the nucleus
sequence number. $\chi^{2}$ is the average values of the $N$ nuclei
with $N$ denoting the numbers of nuclei included in global
parameters search and its value is $45$ in the present work.
$\sigma_{el}^{th}(i,j)$ and $\sigma_{el}^{exp}(i,j)$ are the
theoretical and experimental differential cross sections at the
$j$-th angle with the $i$-th incidence energy, respectively. $\Delta
\sigma_{el}^{exp}(i,j)$ is the corresponding experimental data
error. $N_{n,i}$ is the number of angles for the $n$-th nucleus and
the $i$-th incidence energy. $N_{n,el}$ is the number of incident
energy points of elastic scattering angular distribution for the
$n$-th nucleus.

Through minimizing the average $\chi^{2}$ value for the $45$ nuclei
in Table I with the modified code {\small APMN}, we find an optimal
set of global neutron potential parameters, which are given in Eqs.
(8)$-$(15). With the obtained parameters above, we get the average
value of $\chi^2=32.27$ for the $45$ nuclei. Using the parameters of
Koning and Delaroche~\cite{NPA7132312003}, we obtain the average
value of $\chi^2=30.11$ for the same $45$ nuclei. Therefore, our
parameter set has almost the same good global quality as that of
Koning and Delaroche for the global neutron potential.

We use the optical model potential parameters of ours and Koning
\textit{et al} to calculate the $\chi_n^2$ of a single nucleus for
the $45$ nuclei in Table I. In addition, in order to see the
predictive power, we also calculate the $\chi_n^2$ for other $58$
nuclei listed Table II where the incident energy region and
references are also given. The calculated results for all the $103$
nuclei in Table I and Table II are shown in Table III where our
results are denoted by $\chi_{n1}^{2}$ and that of Koning \textit{et
al} are denoted by $\chi_{n2}^{2}$, respectively.

From Table III, we can see that the value of $\chi_{n1}^2$ is
close to that of $\chi_{n2}^2$ for the nuclei in Table I; The
value of $\chi_{n1}^2$ is much less than that of $\chi_{n2}^2$ for
the nuclides Os, Pt, Th, U, and Pu; The value of $\chi_{n1}^2$ is
also close to that of $\chi_{n2}^2$ for the other nuclei. This
means that our new set of the isospin dependent global
neutron-nucleus optical potential parameters can be as equally
good as that of Koning \textit{et al} to reproduce the
experimental data for neutron as projectile with target ranging
from $^{24}$Mg to $^{209}$Bi. However our results are better than
those of Koning \textit{et al} for the actinide. We would like to
point out that the number of parameters of our optical model
potential is significantly less than that of Koning \textit{et
al}.

Some of the elastic scattering angular distributions obtained with
our global optical potential parameters and with those of Koning
\textit{et al} as well as the corresponding experimental data are
plotted in Figs. 1 to 12. The sold lines are the results calculated
with our parameters, the dashed lines are the results with the
parameters of Koning \textit{et al}, and the points represent the
experimental data. The same symbols are used in all figures. The
experimental data and the corresponding theoretical calculation
results in all figures are in the center of mass (C.M.) system. From
these figures, we can see clearly that our theoretical calculations
can reproduce the experimental data as equally well as those of
Koning \textit{et al} in the targets range from $^{24}$Mg to
$^{209}$Bi, except for some energy points of few nuclei.

From Fig. 1 and Fig. 2, it is seen that both of our theoretical
calculations and those of Koning \textit{et al} can not well
reproduce the experimental data for some energy points of targets
$^{40}$Ca and $^{48}$Ca. This is a well-known
problem~\cite{PRC3825891988}\cite{PRC5811181998} for $^{40}$Ca. It
may be due to the fact that both $^{40}$Ca and $^{48}$Ca are double
magic nuclei and the shell effect corrections may be important.
However, both of our work and that of Koning \textit{et al} aim at
constructing global spherical optical model potentials. So the shell
effects are not included in both of the OMPs. In addition, the
effects of giant resonances have been neglected in both theoretical
calculations and including them could improve the
agreement~\cite{PRC243691981}.

From Figs. 3-8, one can see that there exist some obvious
deviations between experimental data and theoretical calculations
with both our OMP parameters and that of Koning \textit{et al} for
nuclei Ba and W. This may be due to the fact that the Ba and W
exist large deformation, and an effective spherical mean field may
no longer provide a totally adequate description of the
neutron-nucleus many body problem~\cite{NPA7132312003}. Both of
the OMPs are based on spherical frame and the effects of
deformation are not considered.

For the actinide, such as Th, U, and Pu, it is seen from Figs.
9-12 that our theoretical results exhibit significantly better
agreement with experimental data than those of Koning \textit{et
al}.

\section{SUMMARY}

A new set of isospin dependent global neutron-nucleus optical
potential parameters has been obtained based on the existing
experimental data of neutron elastic scattering angular
distributions by using the modified code {\small
APMN}~\cite{NuclSciEng1411782002}. The calculated elastic scattering
angular distributions with the new optical model potential
parameters have been shown to be in good agreement with the
corresponding experimental data for many nuclei from $^{24}$Mg to
$^{242}$Pu in the energy region up to 200 MeV. In particular, our
new global optical model potential parameters can give a
significantly improved description of neutron elastic scattering
angular distributions for the actinide, such as Th, U, and Pu, than
the existing global optical model potential parameters in the
literature. Our new global optical model potential can be used to
calculate the neutron elastic scattering for different target nuclei
including those for which the experimental data are unavailable so
far.

In the present work, polarization of the projectile is not
considered. The polarized neutron beams may play a very important
role in nuclear reaction and nuclear structure studies as well as
many fundamental issues of particle physics. We plan to investigate
the effect of neutron polarization in a future work.

\begin{acknowledgments}
The authors would like to thank Professor Chong-Hai Cai for useful
discussions. This work was supported in part by the NNSF of China
under Grant Nos. 10975097 and 11047157, Shanghai Rising-Star
Program under Grant No. 11QH1401100, and the National Basic
Research Program of China (973 Program) under Contract No.
2007CB815004.
\end{acknowledgments}



\pagebreak
\begin{table}[!htb]
\caption{The data base for searching global optical potential
parameters}
\begin{tabular}{cccccc}
\hline
  nucleus & En.(MeV) & Refs. & nucleus & En.(MeV) & Refs. \\
  \hline
$^{24}$Mg&1.5-14.83&\cite{UFZ1317811968}-\cite{Con7691982} &$^{27}$Al & 0.3-26 &\cite{UFZ1317811968}\cite{UFZ813891963}-\cite{PR1252761962}\\
$^{28}$Si&3.4-21.7&\cite{KE201741977}\cite{Con7691982}\cite{NPA1135641968}\cite{Con1931988}\cite{ComRenB26217361966}-\cite{NPA4585021986}&$^{31}$P & 3.4-20 & \cite{KE201741977}\cite{Con1751961}\cite{NPA1135641968}\cite{NPA196651972}\cite{Con11841978}\cite{NP683691965}\\
$^{32}$S&5.95-21.7&\cite{Con15091965}\cite{Con7691982}\cite{NPA1135641968}\cite{NPA4585021986}\cite{NPA260951976}-\cite{NPA283231977}& $^{40}$Ca  & 1.175-225 &\cite{PREANDC1491971}-\cite{PRC700546132004}\\
$^{45}$Sc&1.6-10&\cite{JPG196551993}&$^{51}$V& 1.61-14.37 &\cite{NPA1176571968}-\cite{Con2091988}\\
$^{52}$Cr&1.5-18.54 &\cite{UFZ228661977}-\cite{PRNEANDC155201990}&$^{55}$Mn&2.47-14.1&\cite{KE201741977}\cite{NPA1501051970}\cite{NP891541966}\cite{Con2091988}\cite{NPA2753251977}\\
$^{54}$Fe&1.3-26&\cite{ANE136011986}-\cite{PRC3820521988}&$^{56}$Fe&1.8-26&\cite{KE201741977}\cite{NPA1615931971}\cite{NPA3905091982}-\cite{Con3621973}\cite{YFIR1481972}-\cite{PRNEANDC155951990} \\
$^{59}$Co&1-23&\cite{NPA1501051970}\cite{PRC680446102003}\cite{NPA4722371987}\cite{NPA2753251977}\cite{PR9310621954}-\cite{PRC32761985}&$^{58}$Ni&1.42-24&\cite{UFZ228661977}\cite{Con3621973}\cite{PRC3820521988}\cite{ZPA3062651982}-\cite{Con3751980}\\
$^{60}$Ni&1.5-24&\cite{UFZ228661977}\cite{Con3621973}\cite{YF31131980}-\cite{ORNL45231970}&$^{63}$Cu&5.5-13.92&\cite{NPA3905091982}\cite{ORNL49081974}\\
$^{65}$Cu&2.33-13.92&\cite{NPA3905091982}\cite{PRCOO157321967}\cite{ORNL49081974}&$^{75}$As&3.2-8.05&\cite{NP891541966}\cite{RAE4821973}\cite{RAE4851974}\\
$^{80}$Se&0.34-10&\cite{Con32131987}-\cite{NPA4202371984}&$^{88}$Sr&11&\cite{NPA3114921978}\\
$^{89}$Y&0.8892-21.6&\cite{PRC311111985}\cite{NPA4722371987}\cite{RANL79351972}-\cite{PRC348251986}&$^{90}$Zr&1.5-24&\cite{PRC1010871974}-\cite{NPA5173011990}\\
$^{91}$Zr&8-24&\cite{NPA5173011990}&$^{92}$Zr&1.5-24&\cite{PRC1010871974}\cite{PRC1217971975}\cite{PAKTY74341977}\cite{NEANDC51111977}\cite{NPA5173011990}\\
$^{94}$Zr&1.5-24&\cite{PRC1010871974}\cite{NPA5173011990}&$^{93}$Nb& 1-20&\cite{PRC311111985}\cite{NPA2753251977}\cite{PR9310621954}\cite{PANL721031966}-\cite{PC1991}\\
$^{92}$Mo&0.9-26&\cite{PRC1010871974}\cite{PAKTY74341977}\cite{NPA20111973}-\cite{NPA31311979}&$^{96}$Mo&0.9-26&\cite{PRC1010871974}\cite{PAKTY74341977}\cite{NPA20111973}\cite{NPA2442131975}\cite{PRC96701974}-\cite{Con13281976}\\
$^{98}$Mo&0.9-26&\cite{NPA20111973}\cite{NPA2442131975}\cite{NPA31311979}&$^{100}$Mo&0.9-26&\cite{PAKTY74341977}\cite{NPA20111973}\cite{NPA2442131975}\cite{PRC96701974}\cite{NPA31311979}\\
$^{103}$Rh&1.5-9.995&\cite{NPA41511984}\cite{NP625111965}\cite{JPG207951994}&$^{107}$Ag&1.5-4&\cite{NPA3322971979}\\
$^{116}$Sn&0.4-24&\cite{YF5015311990}-\cite{NPA341561980}&$^{118}$Sn&0.8-24&\cite{YF5015311990}\cite{NPA341561980}-\cite{NST255111988}\\
$^{120}$Sn&0.4-16.905&\cite{NPA2753251977}\cite{YF5015311990}-\cite{NPA341561980}\cite{Con1481972}&$^{124}$Sn&0.4-24&\cite{YF5015311990}\cite{NPA341561980}\\
$^{127}$I&0.8893-16.1&\cite{DOK1585741964}\cite{ANL79351972}-\cite{NPA3323491979}&$^{141}$Pr&0.8788-8&\cite{NP891541966}\cite{ANL79351972}\cite{NP42861963}-\cite{Con27551968}\\
$^{142}$Nd&2.5-7&\cite{BAP248541979}\cite{PRC20781979}&$^{144}$Nd&2.5-7&\cite{BAP248541979}\cite{PRC20781979}\\
$^{148}$Sm&2.47-7&\cite{PRC1622231977}-\cite{PRC159271977}&$^{197}$Au&0.134-14.7&\cite{PRC311111985}\cite{AE4821973}\cite{ZPA1961031966}\cite{YK1994151994}-\cite{AE4851974}\\
$^{206}$Pb & 0.5-21.6 &\cite{NPA4722371987}\cite{NPA2753251977}\cite{ZPA1961031966}\cite{PC1965}\cite{AE4851974}-\cite{YF156621972}&$^{208}$Pb&1.285-225&\cite{PRC311111985}\cite{PRC700546132004}\cite{YF156621972}-\cite{DAB4057241980}\\
$^{209}$Bi& 2-24&\cite{NPA4722371987}\cite{NPA4432491985}\cite{NSE75691980}-\cite{PR12812711962}&&&\\

\hline \hline
\end{tabular}
\end{table}

\pagebreak
\begin{table}[!htb]
\caption{The incident energy points and data references of the
other 58 nuclei.}
\begin{tabular}{cccccc}
\hline
  nucleus & En.(MeV) & Refs. & nucleus & En.(MeV) & Refs. \\
  \hline
$^{26}$Mg&24&\cite{NPA4012371983}&$^{34}$S & 21.7,25.5 &\cite{NPA4585021986}\\
$^{39}$K&14.07&\cite{ComRenB26217361966}&$^{48}$Ca & 2.35-7.97 &\cite{PHDT1988}\\
$^{48}$K&2.9&\cite{UFZ1418771969}&$^{50}$Cr & 1.5-3 &\cite{UFZ228661977}\\
$^{54}$Cr&1.5-3&\cite{UFZ1418771969}\cite{UFZ191521974}&$^{62}$Ni&1.5-5&\cite{UFZ228661977}\cite{YF31131980}\cite{DOK1585741964}\\
$^{64}$Ni&1.5-7&\cite{UFZ228661977}\cite{YF31131980}&$^{64}$Zn&1.5-3&\cite{UFZ228661977}\cite{UFZ191521974}\\
$^{66}$Zn&1.5-3&\cite{UFZ228661977}\cite{UFZ191521974}&$^{68}$Zn&1.5-3&\cite{UFZ228661977}\\
$^{76}$Se&0.34-10&\cite{Con32131987}-\cite{Con31671983}\cite{NPA4202371984}\cite{PRC149331976}&$^{78}$Se&0.34-8&\cite{Con32131987}-\cite{Con31671983}\cite{PRC149331976}\\
$^{82}$Se&0.34-10&\cite{Con32131987}-\cite{Con31671983}\cite{PRC149331976}&$^{94}$Mo&0.9-8.04&\cite{PRC1010871974}\cite{PAKTY74341977}\cite{NPA20111973}\cite{YK50401983}\cite{PRC96701974}\cite{Con13281976}\\
$^{110}$Cd&0.4-1.24&\cite{YF5015311990}&$^{114}$Cd&4&\cite{DOK1585741964}\\
$^{116}$Cd&0.6-1.24&\cite{YF5015311990}&$^{113}$In&5.19-8.53&\cite{AE512441981}\\
$^{115}$In&1.8-8.53&\cite{DOK1585741964}\cite{AE512441981}\cite{Con12671972}&$^{122}$Sn&0.4-11&\cite{YF5015311990}\cite{NPA341561980}\\
$^{122}$Te&0.3-1.97&\cite{KSF6121987}&$^{124}$Te&0.3-1.97&\cite{KSF6121987}\\
$^{126}$Te&0.3-1.97&\cite{KSF6121987}&$^{128}$Te&0.3-1.97&\cite{KSF6121987}\\
$^{130}$Te&0.3-1.97&\cite{KSF6121987}&$^{133}$Cs&0.8772&\cite{ANL79351972}\\
$^{134}$Ba&3-20&\cite{NSE653681978}&$^{135}$Ba&2-20&\cite{NSE653681978}\\
$^{136}$Ba&4-20&\cite{NSE653681978}&$^{137}$Ba&3-20&\cite{NSE653681978}\\
$^{138}$Ba&5-20&\cite{NSE653681978}&$^{137}$Ba&3-20&\cite{NSE653681978}\\
$^{138}$Ba&5-20&\cite{NSE653681978}&$^{139}$La&0.98-8&\cite{NP42861963}\cite{Con27551968}\\
$^{140}$Ce&7.5-14.6&\cite{PRC311111985}\cite{PRC630146062001}&$^{139}$La&7.5&\cite{PRC630146062001}\\
$^{146}$Nd&2.5-7&\cite{BAP248541979}\cite{PRC20781979}&$^{148}$Nd&2.5-7&\cite{BAP248541979}\cite{PRC20781979}\\
$^{150}$Nd&2.5-7&\cite{BAP248541979}\cite{PRC20781979}&$^{150}$Sm&2.47-7&\cite{PRC1622231977}\cite{PRC159271977}\\
$^{152}$Sm&2.5-7&\cite{PRC1622231977}\cite{PRC159271977}&$^{154}$Sm&6.25-7&\cite{PLB582931975}\cite{PRC159271977}\\
$^{152}$Sm&2.5-7&\cite{PRC1622231977}\cite{PRC159271977}&$^{181}$Ta&0.323-14.8&\cite{PRC311111985}\cite{NPA2753251977}\cite{INDC1181989}\cite{ANE3219262005}-\cite{NPA2121471973}\\
$^{182}$W&1.5-4.87&\cite{PRC2624331982}\cite{NPA4422341985}&$^{184}$W&1.5-4.84&\cite{PRC2624331982}\cite{NPA4422341985}\\
$^{182}$W&1.5-4.87&\cite{PRC2624331982}\cite{NPA4422341985}&$^{184}$W&1.5-4.84&\cite{PRC2624331982}\cite{NPA4422341985}\\
$^{186}$W&1.5-3.95&\cite{PRC2624331982}&$^{190}$Os&2.5-4&\cite{PHDT1987}\\
$^{192}$Os&1.6-3.94&\cite{PRC4025091989}&$^{194}$Pt&2.5-4.55&\cite{PRC3214881985}\cite{PRC36731987}\\
$^{196}$Pt&2.53-4.64&\cite{PHDT1987}&$^{204}$Pb&2.53-8&\cite{PRC491031994}\\
$^{207}$Pb&0.5-13.7&\cite{PC1965}\cite{NSE651741978}\cite{YF156621972}\cite{EANDC3061973}&$^{232}$Th&0.144-14.1&\cite{PR9310621954}\cite{NSE814911982}\cite{NST209831983}-\cite{PRC3420751986}\\
$^{233}$U&0.7-1.5&\cite{NSE814911982}&$^{235}$U&0.185-5.5&\cite{NSE814911982}\cite{PR1047311956}-\cite{RAWRE551969}\cite{PHDT1986}\\
$^{238}$U&0.055-15&\cite{NSE814911982}\cite{NP652361965}\cite{AE621921987}-\cite{AE113951961}&$^{239}$Pu&0.149-14.1&\cite{NSE814911982}\cite{PRC3420751986}\cite{PR1047311956}\cite{RAERE59721969}\cite{ZPA2322861970}\\
$^{240}$Pu&0.4-1.2&\cite{NSE47191972}&$^{242}$Pu&0.57-1.5&\cite{RLA78551979}\\

\hline \hline
\end{tabular}
\end{table}

\pagebreak
\begin{table}
\caption{\label{tab:table2}$\chi_{n}^2$ of a single nucleus.
$\chi_{n1}^2$ for our global potential parameters, $\chi_{n2}^2$ for
those of A. J. Koning and J. P. Delaroche}
\begin{ruledtabular}
\renewcommand{\arraystretch}{0.8}
\begin{tabular}{ccccccccc}
nucleus&$\chi_{n1}^2$&$\chi_{n2}^2$&nucleus&$\chi_{n1}^2$&$\chi_{n2}^2$&nucleus&$\chi_{n1}^2$&$\chi_{n2}^2$\\
\hline
$^{24}$Mg  & 48.07 & 77.88 & $^{26}$Mg  & 73.45  & 43.48 & $^{27}$Al  & 33.74  & 50.36 \\
$^{28}$Si  & 27.24 & 22.88 & $^{31}$P   & 33.46  & 43.17  & $^{32}$S  & 13.25  & 14.51 \\
$^{34}$S   & 18.78 & 15.66 & $^{39}$K   & 65.04  & 45.88 & $^{40}$Ca  & 19.44  & 16.25 \\
$^{48}$Ca  & 319.3 & 301.2 & $^{45}$Sc  & 14.08  & 7.934 & $^{48}$Ti  & 16.45  & 12.60 \\
$^{51}$V   & 29.21 & 21.82 & $^{50}$Cr  & 3.095  & 5.057 & $^{52}$Cr  & 11.81  & 94.89 \\
$^{54}$Cr  & 3.012 & 4.558 & $^{55}$Mn  & 17.92  & 23.83 & $^{54}$Fe  & 31.20  & 122.7 \\
$^{56}$Fe  & 29.40 & 38.53 & $^{59}$Co  & 50.25  & 51.38 & $^{58}$Ni  & 11.64  & 20.90 \\
$^{60}$Ni  & 18.90 & 27.08 & $^{62}$Ni  & 6.552  & 9.851 & $^{64}$Ni  & 6.625  & 5.580 \\
$^{63}$Cu  & 8.719 & 6.573 & $^{65}$Cu  & 10.11  & 5.919 & $^{64}$Zn  & 6.833  & 7.435 \\
$^{66}$Zn  & 6.611 & 4.363 & $^{68}$Zn  & 7.712  & 3.665 & $^{75}$As  & 12.06  & 11.64 \\
$^{76}$Se  & 44.21 & 63.26 & $^{78}$Se  & 22.08  & 35.32 & $^{80}$Se  & 29.91  & 39.48 \\
$^{82}$Se  & 13.97 & 14.68 & $^{88}$Sr  & 38.11  & 32.66 & $^{89}$Y   & 34.55  & 19.09 \\
$^{90}$Zr  & 29.93 & 23.97 & $^{91}$Zr  & 31.67  & 19.35 & $^{92}$Zr  & 13.97  & 6.971 \\
$^{94}$Zr  & 13.06 & 12.79 & $^{93}$Nb  & 50.10  & 42.60 & $^{92}$Mo  & 24.39  & 29.55 \\
$^{94}$Mo  & 40.40 & 35.57 & $^{96}$Mo  & 101.8  & 170.6 & $^{98}$Mo  & 17.66  & 20.94 \\
$^{100}$Mo & 67.80 & 137.1 & $^{103}$Rh & 10.33  & 15.54 & $^{107}$Ag & 8.043  & 42.19 \\
$^{110}$Cd & 4.622 & 6.179 & $^{114}$Cd & 298.4  & 77.41 & $^{113}$In & 5.413  & 5.741 \\
$^{115}$In & 51.65 & 18.74 & $^{116}$Sn & 8.380  & 8.194 & $^{118}$Sn & 11.67  & 13.38 \\
$^{120}$Sn & 28.10 & 17.12 & $^{122}$Sn & 9.091  & 4.693 & $^{124}$Sn & 10.06  & 5.025 \\
$^{122}$Te & 3.090 & 8.610 & $^{124}$Te & 2.177  & 5.716 & $^{126}$Te & 3.471  & 4.347 \\
$^{128}$Te & 9.282 & 3.668 & $^{130}$Te & 13.68  & 4.837 & $^{127}$I  & 91.22  & 50.50 \\
$^{133}$Cs & 7.464 & 7.211 & $^{134}$Ba & 897.9  & 726.0 & $^{135}$Ba & 687.2  & 292.9 \\
$^{136}$Ba & 957.7 & 300.9 & $^{137}$Ba & 1221.  & 378.3 & $^{138}$Ba & 2055.  & 457.7 \\
$^{139}$La & 44.88 & 41.60 & $^{140}$Ce & 44.40  & 13.34 & $^{142}$Ce & 199.8  & 129.7 \\
$^{141}$Pr & 125.2 & 115.1 & $^{142}$Nd & 27.00  & 20.94 & $^{144}$Nd & 12.55  & 8.944 \\
$^{146}$Nd & 14.58 & 22.04 & $^{148}$Nd & 23.58  & 96.56 & $^{150}$Nd & 137.8  & 319.3 \\
$^{148}$Sm & 18.90 & 30.05 & $^{150}$Sm & 18.01  & 87.68 & $^{152}$Sm & 29.60  & 113.2 \\
$^{154}$Sm & 34.01 & 28.10 & $^{181}$Ta & 33.85  & 92.52 & $^{182}$W  & 75.12  & 316.4 \\
$^{184}$W  & 65.26 & 252.6 & $^{186}$W  & 65.85  & 234.4 & $^{190}$Os & 246.5  & 636.7 \\
$^{192}$Os & 96.69 & 323.6 & $^{194}$Pt & 78.25  & 297.1 & $^{196}$Pt & 69.51  & 204.0 \\
$^{197}$Au & 53.09 & 42.39 & $^{204}$Pb & 48.68  & 48.38 & $^{206}$Pb & 48.26  & 34.32 \\
$^{207}$Pb & 41.30 & 9.177 & $^{208}$Pb & 39.23  & 26.93 & $^{209}$Bi & 50.53  & 24.29 \\
$^{232}$Th & 43.23 & 293.5 & $^{233}$U  & 77.01  & 241.0 & $^{235}$U  & 33.50  & 126.2 \\
$^{238}$U  & 119.3 & 551.8 & $^{239}$Pu & 37.75  & 143.5 & $^{240}$Pu & 36.86  & 175.9 \\
$^{242}$Pu & 27.92 & 91.89 &            &        &       &            &        &       \\
\end{tabular}
\end{ruledtabular}
\end{table}

\begin{figure}[tbp]
\includegraphics[width=18cm,height=20cm]{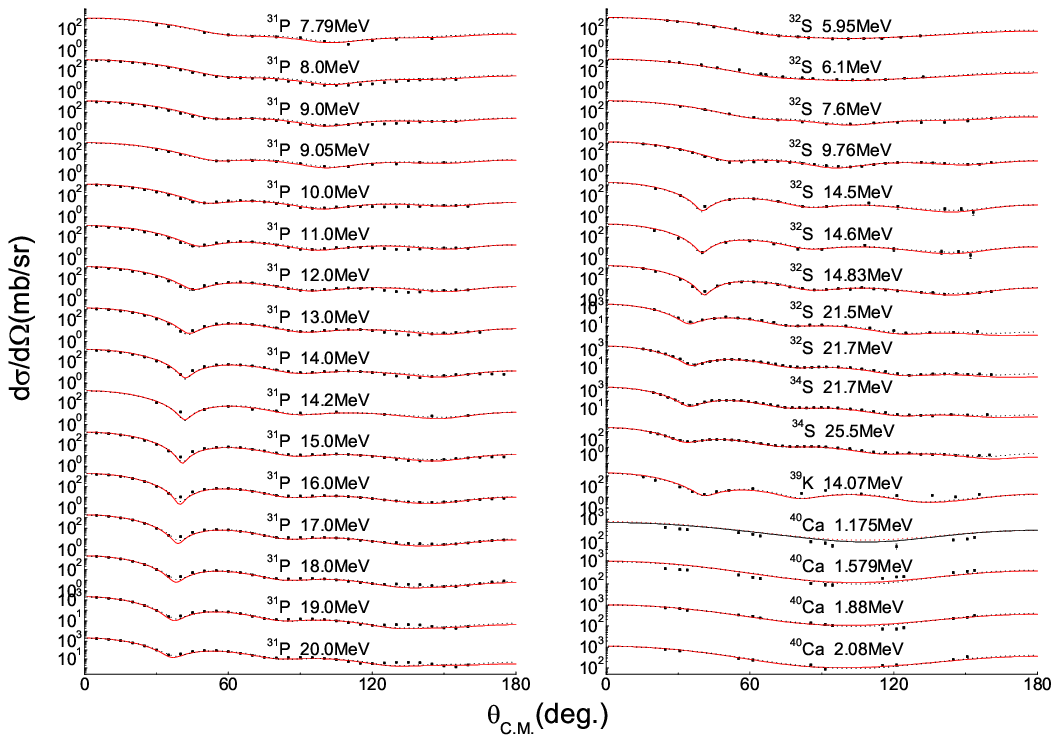}
\caption{Comparisons of the experimental angular distributions of
elastic scattering (dots) with the calculated results from our
global potential parameters (red solid lines) and those of A. J.
Koning and J. P. Delaroche (black dashed lines) in the center of
mass frame. The experimental data are taken from
Refs.~\cite{Con15091965}\cite{Con7691982}\cite{NPA1135641968}\cite{NPA196651972}\cite{ComRenB26217361966}\cite{NPA4585021986}-\cite{PREANDC1491971}.}
\end{figure}

\begin{figure}[tbp]
\includegraphics[width=18cm,height=20cm]{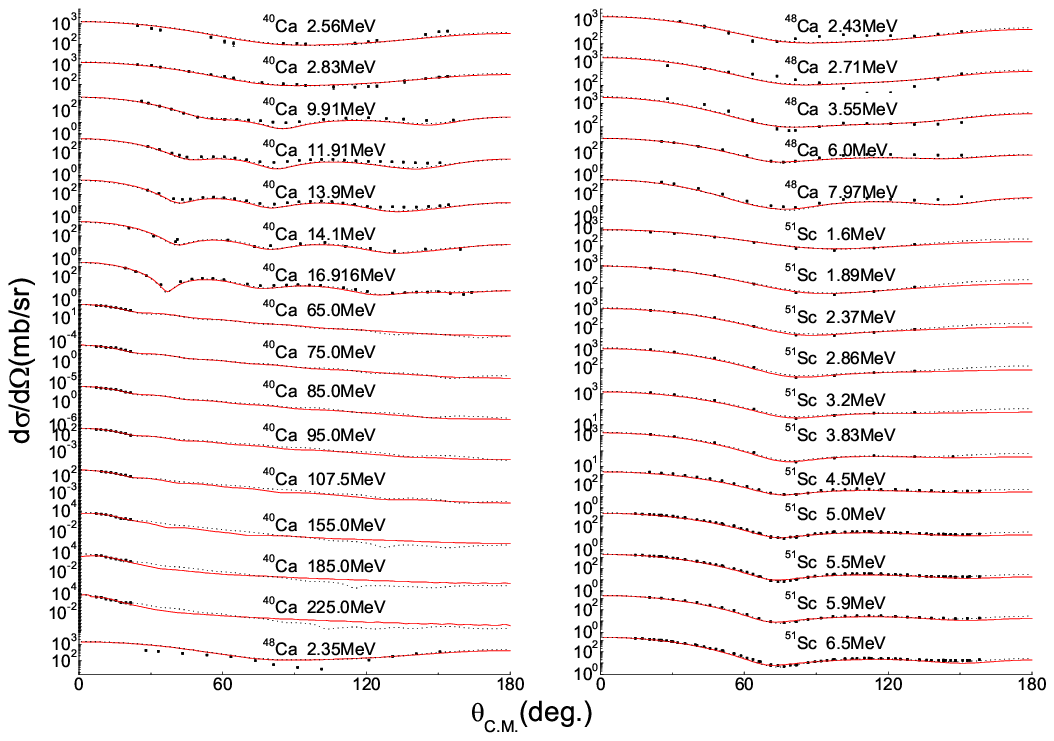}
\caption{Comparisons of the experimental angular distributions of
elastic scattering (dots) with the calculated results from our
global potential parameters (red solid lines) and those of A. J.
Koning and J. P. Delaroche (black dashed lines) in the center of
mass frame. The experimental data are taken from
Refs.~\cite{PREANDC1491971}-\cite{JPG196551993}\cite{PHDT1988}\cite{PRC4125601990}.}
\end{figure}

\begin{figure}[tbp]
\includegraphics[width=18cm,height=20cm]{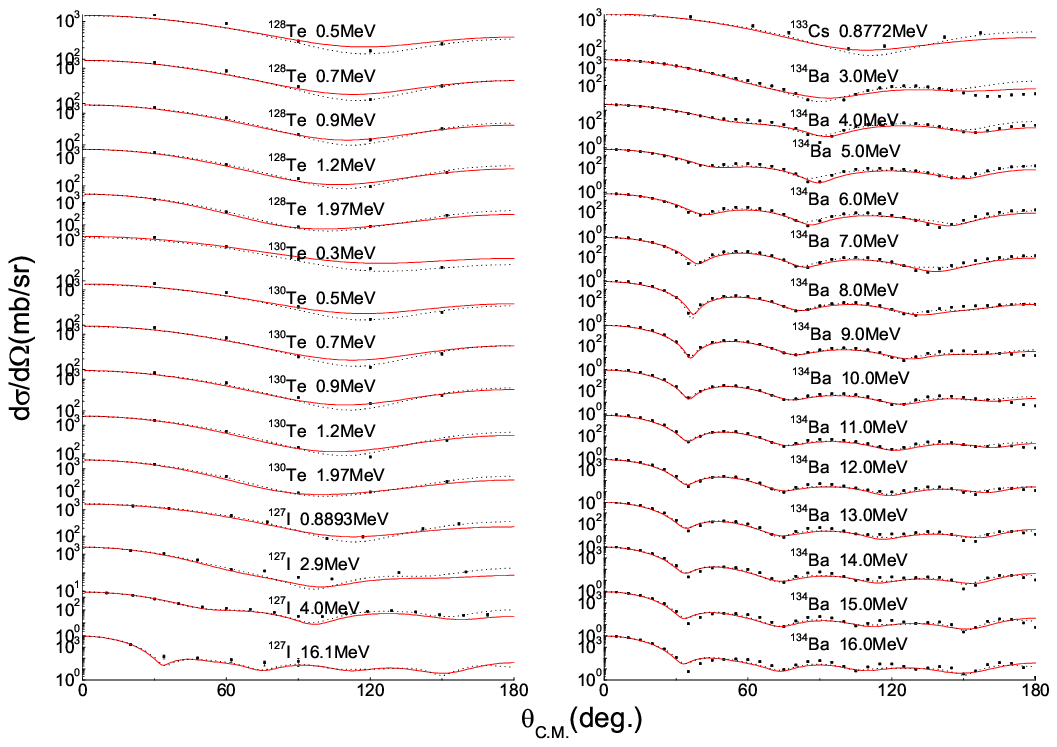}
\caption{Comparisons of the experimental angular distributions of
elastic scattering (dots) with the calculated results from our
global potential parameters (red solid lines) and those of A. J.
Koning and J. P. Delaroche (black dashed lines) in the center of
mass frame. The experimental data are taken from
Refs.~\cite{DOK1585741964}\cite{ANL79351972}-\cite{NPA3323491979}\cite{KSF6121987}\cite{NSE653681978}.}
\end{figure}
\begin{figure}[tbp]
\includegraphics[width=18cm,height=20cm]{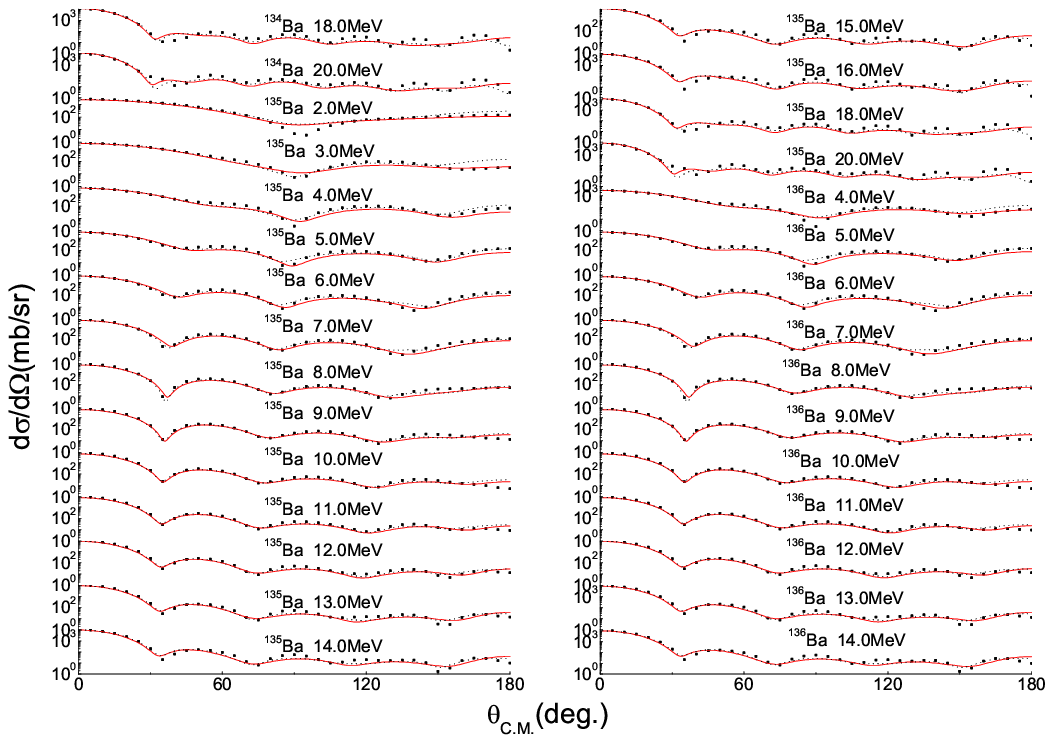}
\caption{Comparisons of the experimental angular distributions of
elastic scattering (dots) with the calculated results from our
global potential parameters (red solid lines) and those of A. J.
Koning and J. P. Delaroche (black dashed lines) in the center of
mass frame. The experimental data are taken from
Refs.~\cite{KSF6121987}.}
\end{figure}
\begin{figure}[tbp]
\includegraphics[width=18cm,height=20cm]{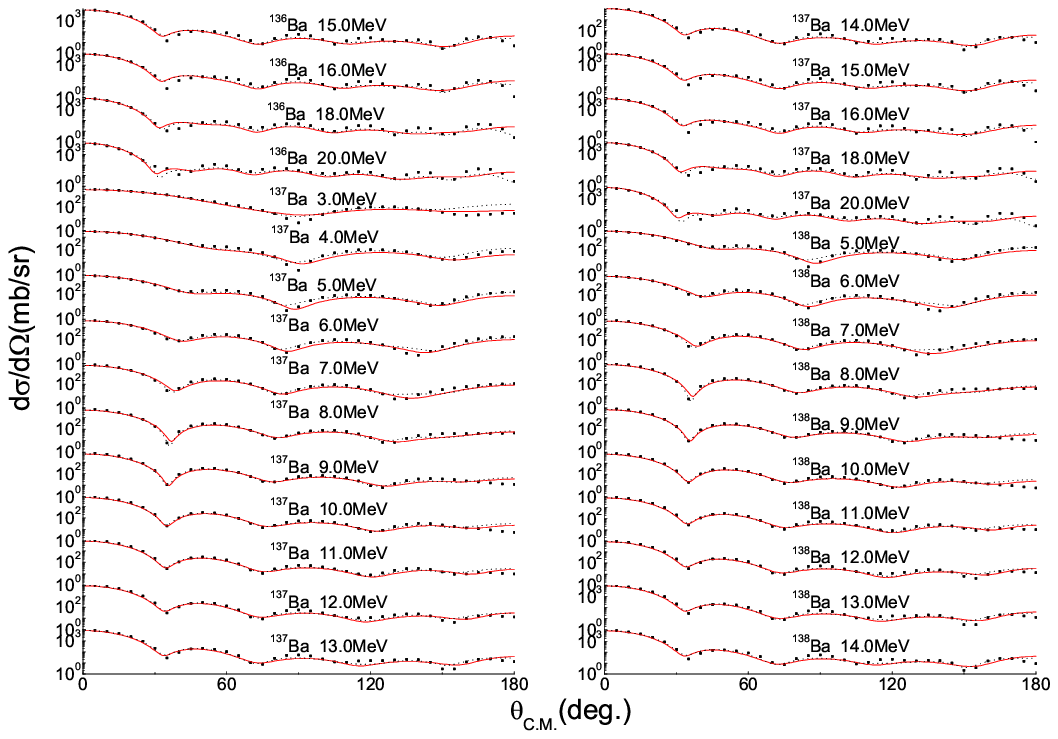}
\caption{Comparisons of the experimental angular distributions of
elastic scattering (dots) with the calculated results from our
global potential parameters (red solid lines) and those of A. J.
Koning and J. P. Delaroche (black dashed lines) in the center of
mass frame. The experimental data are taken from
Refs.~\cite{KSF6121987}.}
\end{figure}
\begin{figure}[tbp]
\includegraphics[width=18cm,height=20cm]{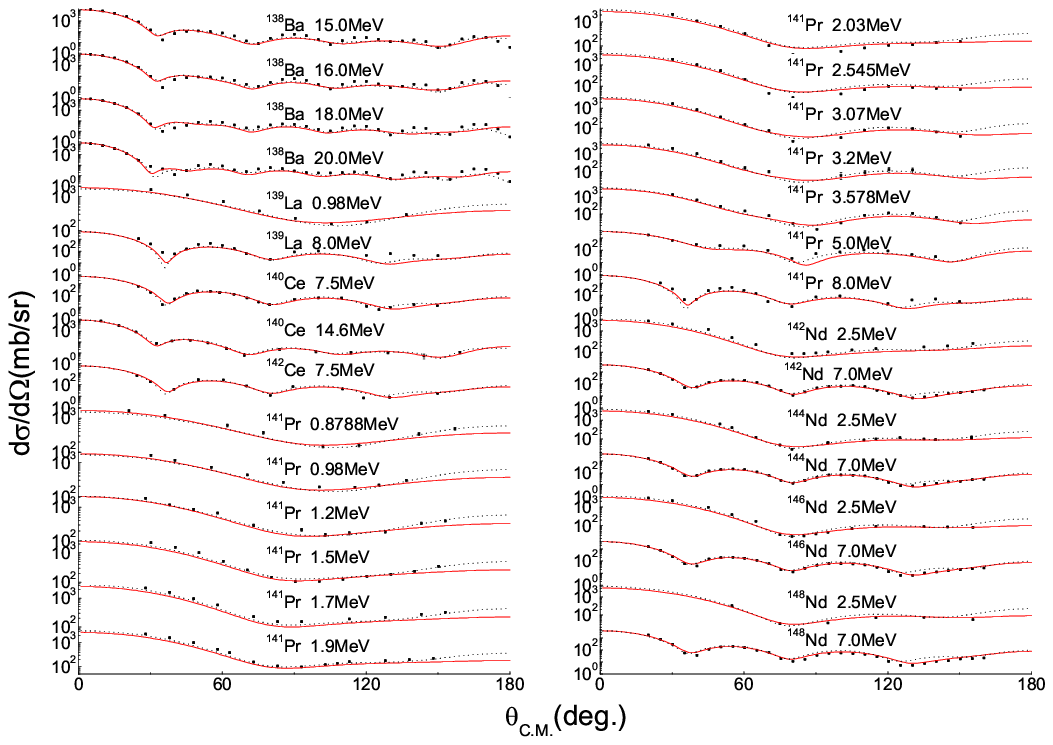}
\caption{Comparisons of the experimental angular distributions of
elastic scattering (dots) with the calculated results from our
global potential parameters (red solid lines) and those of A. J.
Koning and J. P. Delaroche (black dashed lines) in the center of
mass frame. The experimental data are taken from
Refs.~\cite{NP891541966}\cite{PRC311111985}\cite{ANL79351972}\cite{NP42861963}-\cite{PRC20781979}\cite{NSE653681978}\cite{PRC630146062001}.}
\end{figure}
\begin{figure}[tbp]
\includegraphics[width=18cm,height=20cm]{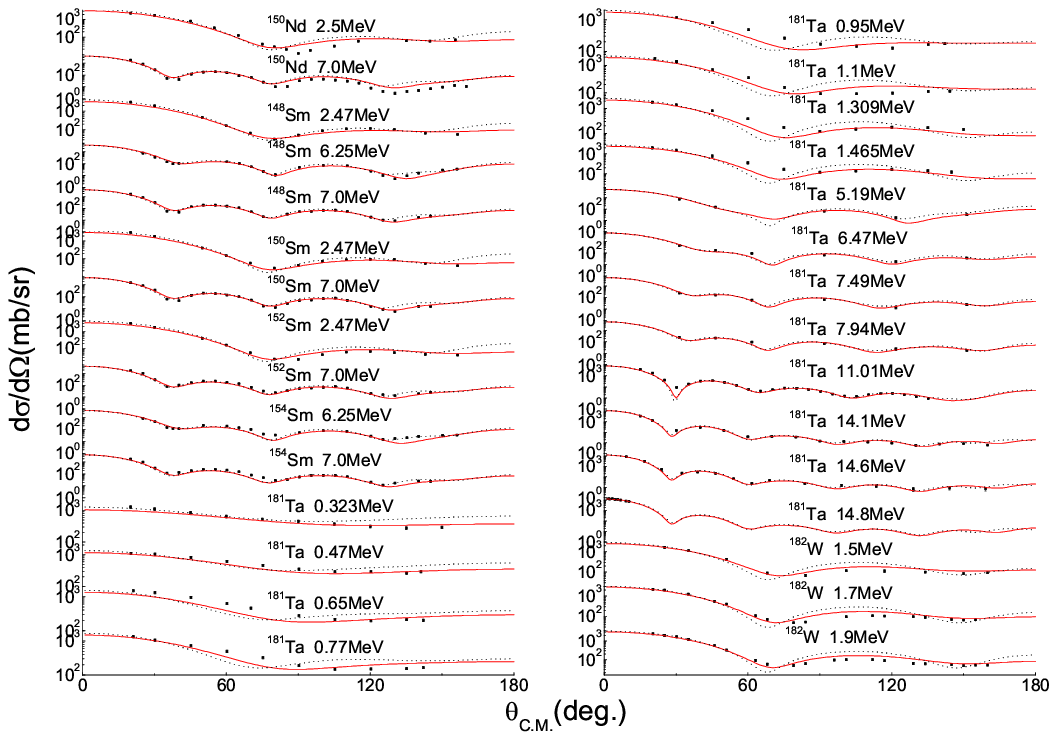}
\caption{Comparisons of the experimental angular distributions of
elastic scattering (dots) with the calculated results from our
global potential parameters (red solid lines) and those of A. J.
Koning and J. P. Delaroche (black dashed lines) in the center of
mass frame. The experimental data are taken from
Refs.~\cite{PRC311111985}\cite{NPA2753251977}\cite{INDC1181989}\cite{BAP248541979}-\cite{PRC159271977}\cite{ANE3219262005}-\cite{PRC2624331982}.}
\end{figure}
\begin{figure}[tbp]
\includegraphics[width=18cm,height=20cm]{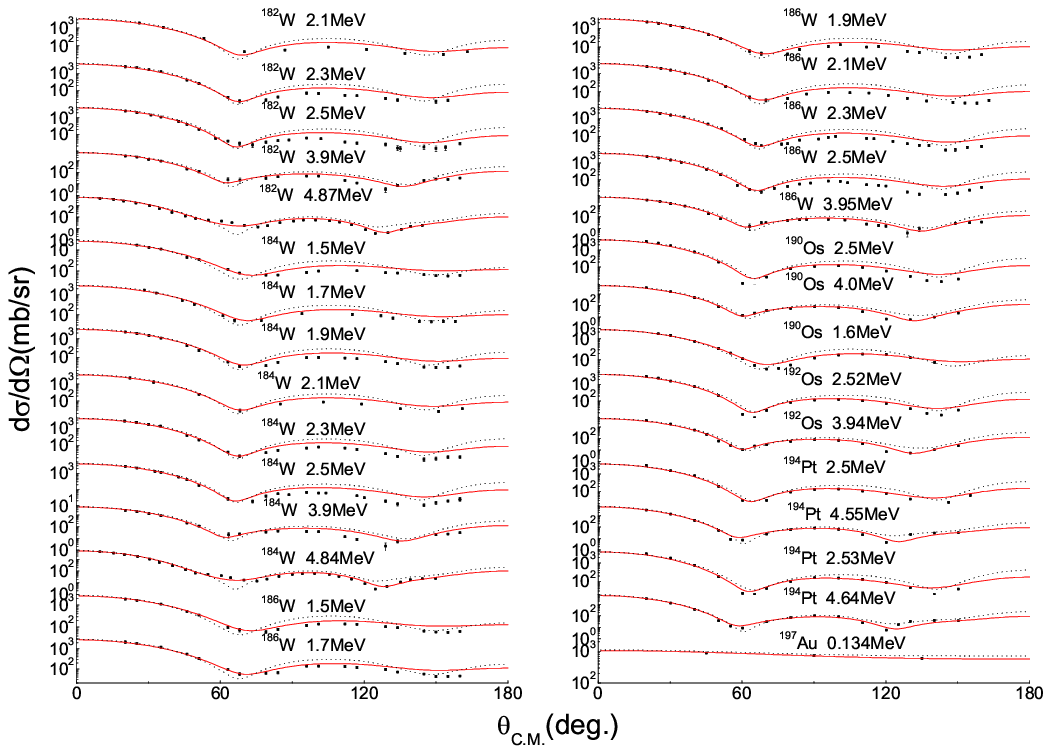}
\caption{Comparisons of the experimental angular distributions of
elastic scattering (dots) with the calculated results from our
global potential parameters (red solid lines) and those of A. J.
Koning and J. P. Delaroche (black dashed lines) in the center of
mass frame. The experimental data are taken from
Refs.~\cite{YK1994151994}\cite{PC1965}\cite{PRC2624331982}-\cite{PRC36731987}.}
\end{figure}

\begin{figure}[tbp]
\includegraphics[width=18cm,height=20cm]{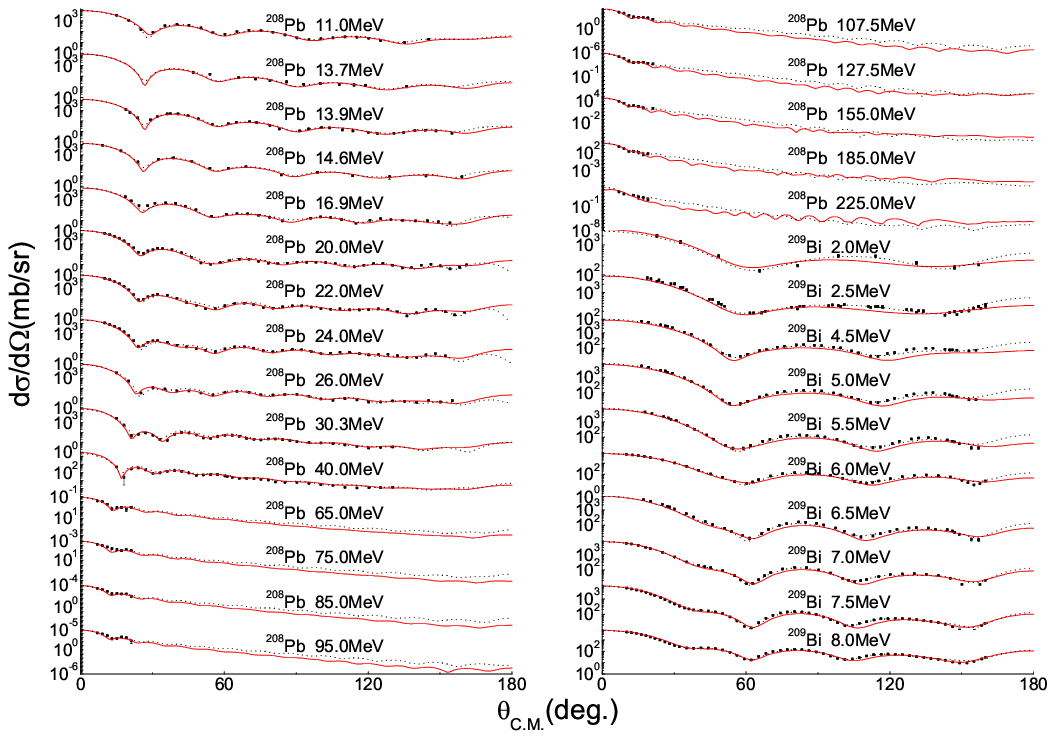}
\caption{Comparisons of the experimental angular distributions of
elastic scattering (dots) with the calculated results from our
global potential parameters (red solid lines) and those of A. J.
Koning and J. P. Delaroche (black dashed lines) in the center of
mass frame. The experimental data are taken from
Refs.~\cite{PRC311111985}\cite{PRC700546132004}\cite{YF156621972}\cite{NPA4432491985}\cite{NPA296951978}\cite{PHDT1981}-\cite{PRC4210131990}.}
\end{figure}
\begin{figure}[tbp]
\includegraphics[width=18cm,height=20cm]{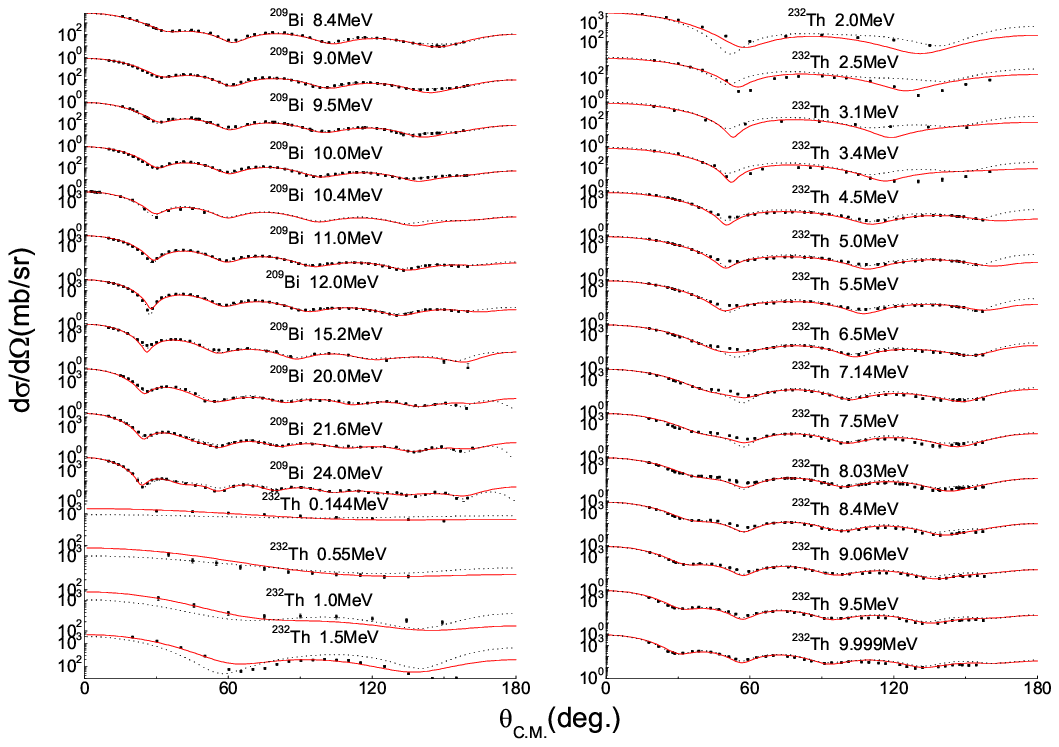}
\caption{Comparisons of the experimental angular distributions of
elastic scattering (dots) with the calculated results from our
global potential parameters (red solid lines) and those of A. J.
Koning and J. P. Delaroche (black dashed lines) in the center of
mass frame. The experimental data are taken from
Refs.~\cite{NPA4722371987}\cite{PR9310621954}\cite{NSE814911982}\cite{PRC3612981987}-\cite{PR12812711962}\cite{NST209831983}-\cite{ANE234591996}.}
\end{figure}
\begin{figure}[tbp]
\includegraphics[width=18cm,height=20cm]{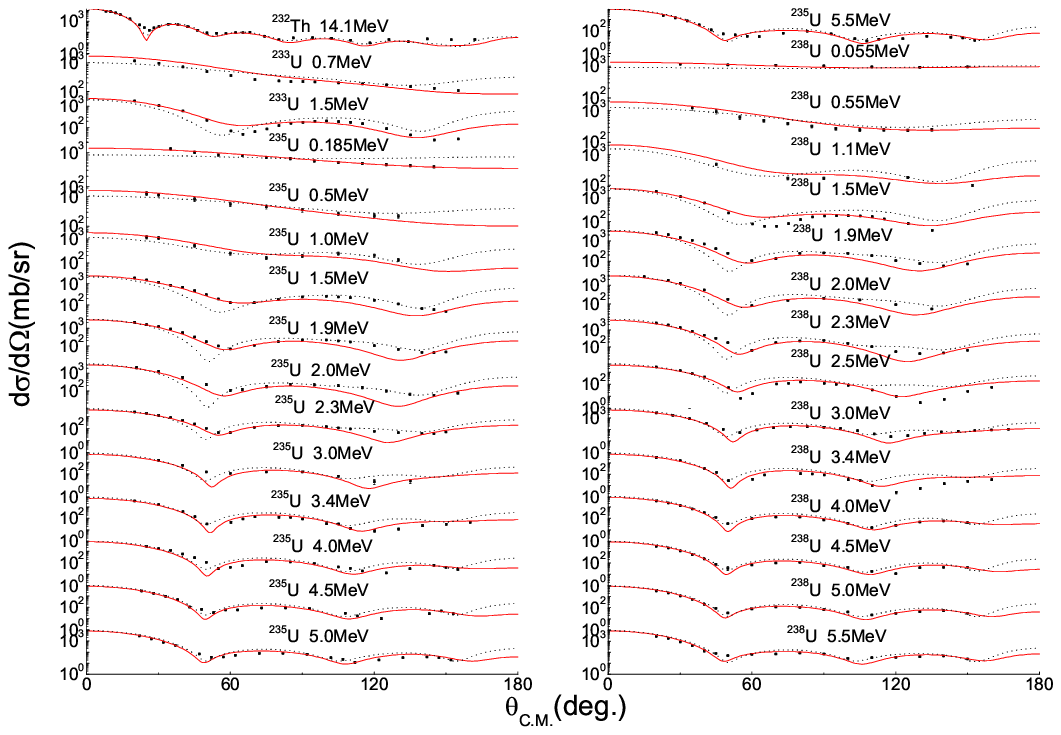}
\caption{Comparisons of the experimental angular distributions of
elastic scattering (dots) with the calculated results from our
global potential parameters (red solid lines) and those of A. J.
Koning and J. P. Delaroche (black dashed lines) in the center of
mass frame. The experimental data are taken from
Refs.~\cite{NSE814911982}\cite{NP652361965}\cite{PRC3420751986}-\cite{JPG1113411985}.}
\end{figure}
\begin{figure}[tbp]
\includegraphics[width=18cm,height=20cm]{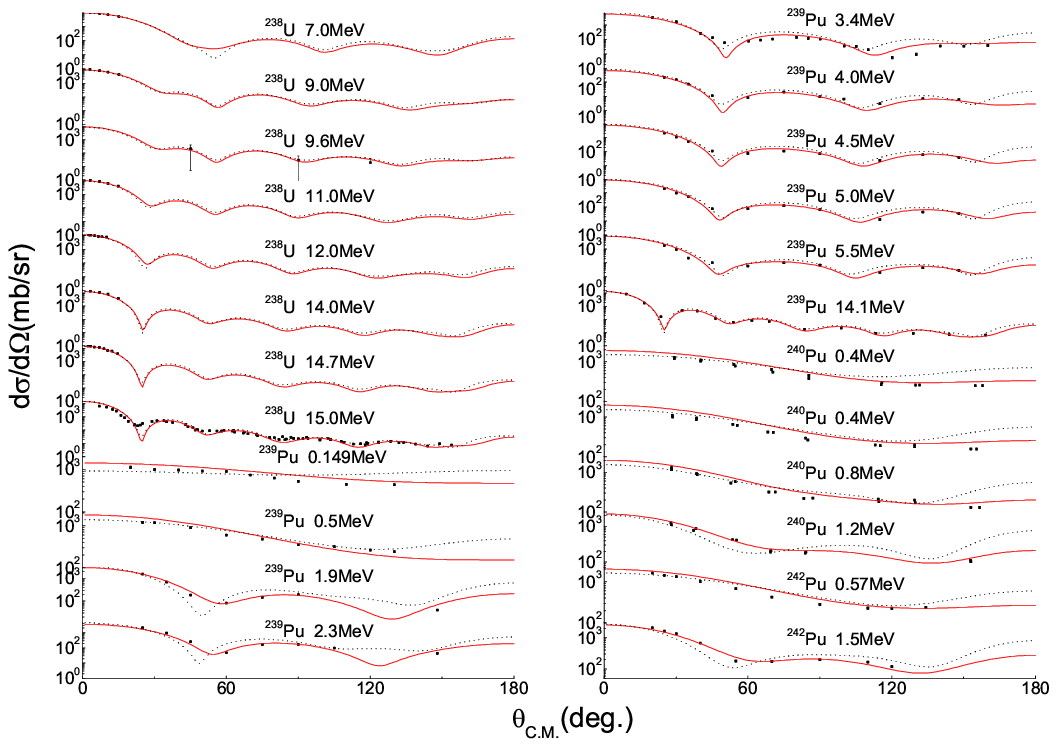}
\caption{Comparisons of the experimental angular distributions of
elastic scattering (dots) with the calculated results from our
global potential parameters (red solid lines) and those of A. J.
Koning and J. P. Delaroche (black dashed lines) in the center of
mass frame. The experimental data are taken from
Refs.~\cite{NSE814911982}\cite{PRC3420751986}\cite{PR1047311956}\cite{PRL3514191975}-\cite{RLA78551979}.}
\end{figure}


\end{document}